\documentclass{emulateapj}

\usepackage{graphicx}
\usepackage{natbib}

\shorttitle{Detailed Radio View on Two Stellar Explosions}
\shortauthors{van der Horst et al.}

\begin{document}

\title{Detailed Radio View on Two Stellar Explosions and Their Host Galaxy: XRF\,080109\,/\,SN\,2008D and SN\,2007uy in NGC\,2770}

\author{A.~J.~van~der~Horst\altaffilmark{1}, 
A.~P.~Kamble\altaffilmark{2}, 
Z.~Paragi\altaffilmark{3,4}, 
L.~J.~Sage\altaffilmark{5}, 
S.~Pal\altaffilmark{6}, 
G.~B.~Taylor\altaffilmark{7},
C.~Kouveliotou\altaffilmark{8}, 
J.~Granot\altaffilmark{9}, 
E.~Ramirez-Ruiz\altaffilmark{10}, 
C.~H.~Ishwara-Chandra\altaffilmark{11}, 
T.~A.~Oosterloo\altaffilmark{12,13}, 
R.~A.~M.~J.~Wijers\altaffilmark{2}, 
K.~Wiersema\altaffilmark{14}, 
R.~G.~Strom\altaffilmark{12,2}, 
D.~Bhattacharya\altaffilmark{15}, 
E.~Rol\altaffilmark{2}, 
R.~L.~C.~Starling\altaffilmark{14}, 
P.~A.~Curran\altaffilmark{16}, 
M.~A.~Garrett\altaffilmark{12,17,18}
}

\altaffiltext{1}{NASA Postdoctoral Program Fellow, Space Science Office, NASA/Marshall Space Flight Center, 
	Huntsville, AL 35812; Alexander.J.VanDerHorst@nasa.gov.}
\altaffiltext{2}{Astronomical Institute, University of Amsterdam, 
	Science Park 904, 1098 XH Amsterdam, The Netherlands.}
\altaffiltext{3}{Joint Institute for VLBI in Europe (JIVE), 
	Postbus 2, 7990 AA Dwingeloo, The Netherlands.}
\altaffiltext{4}{MTA Research Group for Physical Geodesy and Geodynamics, 
	H-1585 Budapest, P. O. Box 585, Hungary.}
\altaffiltext{5}{University of Maryland, Department of Astronomy, 
	College Park, MD 20742.}
\altaffiltext{6}{International Centre for Radio Astronomical Research, University of Western Australia, 
	7 Fairway, Crawley, 6009, Australia.}
\altaffiltext{7}{University of New Mexico, Department of Physics and Astronomy, 
	MSC07 4220, 800 Yale Blvd NE Albuquerque, New Mexico 87131-0001. 
	G.B. Taylor is also an Adjunct Astronomer at the National Radio Astronomy Observatory.}
\altaffiltext{8}{Space Science Office, NASA/Marshall Space Flight Center, 
	Huntsville, AL 38512, USA.}
\altaffiltext{9}{Centre for Astrophysics Research, University of Hertfordshire, 
	College Lane, Hatfield, Herts, AL10 9AB, UK.}
\altaffiltext{10}{Department of Astronomy and Astrophysics, 
	University of California, Santa Cruz, CA 95064.}
\altaffiltext{11}{National Centre for Radio Astrophysics, 
	Post Bag 3, Ganeshkind, Pune 411007, India.}
\altaffiltext{12}{Netherlands Institute for Radio Astronomy (ASTRON), 
	Postbus 2, 7990 AA Dwingeloo, The Netherlands.}
\altaffiltext{13}{Kapteyn Astronomical Institute, University of Groningen, 
	Postbus 800, 9700 AV Groningen, the Netherlands.}
\altaffiltext{14}{Department of Physics \& Astronomy, 
	University of Leicester, Leicester, LE1 7RH, UK.}
\altaffiltext{15}{Inter-University Center for Astronomy and Astrophysics, 
	Pune, Ganeshkhind, Post-bag No. 4, India.}
\altaffiltext{16}{Mullard Space Science Laboratory, University College of London, 
	Holmbury St Mary, Dorking, Surrey RH5 6NT, UK.}
\altaffiltext{17}{Leiden Observatory, University of Leiden, 
	P.B. 9513, Leiden 2300 RA, the Netherlands.}
\altaffiltext{18}{Centre for Astrophysics and Supercomputing, 
	Swinburne University of Technology, Hawthorn, Victoria 3122, Australia.}

% Abstract 
\begin{abstract} 
The galaxy NGC\,2770 hosted two core-collapse supernova explosions, SN\,2008D and SN\,2007uy, 
within 10 days of each other and 9 years after the first supernova of the same type, SN\,1999eh, was found in that galaxy. 
In particular SN\,2008D attracted a lot of attention due to the detection of an X-ray outburst, 
which has been hypothesized to be caused by either a (mildly) relativistic jet or the supernova shock breakout. 
We present an extensive study of the radio emission from SN\,2008D and SN\,2007uy: 
flux measurements with the Westerbork Synthesis Radio Telescope and the Giant Metrewave Radio Telescope, 
covering $\sim$600 days with observing frequencies ranging from 325~MHz to 8.4~GHz. 
The results of two epochs of global Very Long Baseline Interferometry observations are also discussed. 
We have examined the molecular gas in the host galaxy NGC\,2770 with the Arizona Radio Observatory 12-m telescope, 
and present the implications of our observations for the star formation and seemingly high SN rate in this galaxy. 
Furthermore, we discuss the near-future observing possibilities of the two SNe and their host galaxy 
at low radio frequencies with the Low Frequency Array.
\end{abstract}


% Introduction 
\section{Introduction} \label{section:intro}

According to the prevailing model, gamma-ray bursts (GRBs) 
of the long duration class \citep[LGRBs;][]{kouveliotou1993apj} and the spectrally soft and less-energetic 
X-ray flashes \citep[XRFs; e.g.][]{heise2001conf} are caused by a relativistic jet emerging from 
the collapse of a massive star, e.g. the collapsar model \citep[][]{woosley1993apj} 
or through the production of a millisecond magnetar \citep[][]{usov1992nature,bucciantini2009mnras}. 
This drives a long-lived relativistic blast wave into the ambient medium, which accelerates relativistic 
electrons that emit synchrotron radiation observed from radio 
to X-ray frequencies $-$ the afterglow \citep[][]{meszaros1997apj,wijers1997mnras}.

Since 1998 a firm association has been established between LGRBs/XRFs 
and core-collapse supernovae \citep[SNe;][]{galama1998nature,hjorth2003nature,stanek2003apjl,malesani2004apjl,pian2006nature,starling2010arxiv,chornock2010arxiv}. 
Up to now, five LGRBs (of which two are XRFs) have been associated with spectroscopically identified type Ic SNe. 
It has, however, been shown that not all LGRBs/XRFs have an associated SN \citep[][]{gehrels2006nature,fynbo2006nature}. 
Their relative event rates suggest that only a small fraction ($\sim 10^{-3}$) of Ic SNe produce highly luminous GRBs \citep[][]{podsiadlowski2004,soderberg2006nature}. 
A detailed study of the first four GRB-SN associations \citep[][]{kaneko2007apj} has
demonstrated that the total energy budget of the SN explosion only moderately
varies between different events, while the fraction of that energy
that ends up in highly relativistic ejecta (GRB) can vary much more
dramatically between different events. 
In fact, only one GRB out of these five GRB-SN associations, GRB\,030329, 
can be considered a member of the high-luminosity long GRB class, 
both in terms of its prompt and afterglow emission, albeit at the lower end of the distributions 
in terms of its isotropic equivalent energy output in gamma rays \citep[e.g.][]{kaneko2007apj}. 
Therefore, a much larger fraction ($\leq 5-10\%$) of type Ic SNe may produce less
relativistic collimated outflows with Lorentz factors of $2 < \Gamma
\ll 100$ that can power XRFs, compared to $\Gamma \geq 100$ jets that
produce bright GRBs \citep[e.g.][]{granot2004apj,soderberg2006nature}. 

\citet[][]{paragi2010nature} argue that it is very well possible that most or even all Ib/c SNe 
produce (mildly) relativistic ejecta, but in most cases these ejecta carry a much smaller fraction 
of the explosion energy than in GRBs/XRFs, making them detectable in the radio only for very nearby events. 
Current observational constraints \citep[e.g.][]{soderberg2006apj} on these radio sources could be satisfied 
by low-energy mildly relativistic jets and/or relatively low values for external density or shock microphysics parameters. 
It has also been proposed \citep[][]{paczynski2001aca,granot2003apj} 
that some Ib/c SNe are producing relativistic GRB/XRF
jets that point away from our line of sight and are thus not detected
at early times in optical and X-rays, but could be detected in radio bands 
after a few months to years (orphan afterglows). 
After initially unsuccessful efforts to find this latter type of afterglow \citep[e.g.][]{soderberg2006apj}, 
there is currently evidence that two of these events, SN\,2007gr \citep[][]{paragi2010nature} 
and SN\,2009bb \citep[][]{soderberg2010nature} have been observed, 
although their radio emission was already detected within days of the initial explosions. 
In the case of the very nearby SN\,2007gr ($d \approx 11\;$Mpc), mildly-relativistic expansion of the radio source 
was measured using Very Long Baseline Interferometry (VLBI) observations, 
while for the radio-luminous SN\,2009bb (at $d \approx 40\;$Mpc) a similar expansion speed of the ejecta was inferred 
from modeling the broadband radio data. 
We note that the inferred energy in (mildly) relativistic ejecta was significantly larger in SN\,2009bb than in SN\,2007gr, 
$\sim 10^{49}$ and $\sim 10^{46}$ erg, respectively.

The discovery of a bright X-ray transient in the nearby ($d \approx 27\;$Mpc) 
galaxy NGC\,2770 \citep[][]{berger2008atel}, which was later identified as a type Ib~SN, 
has provided an unprecedented opportunity to study the sequence: GRBs -- XRFs -- 
normal core-collapse SNe. The very early discovery and close proximity of this source has 
enabled a multitude of observations across the whole electromagnetic spectrum. 
Follow-up observations of XRF\,080109 detected the transient at optical wavelengths 
\citep[][]{deng2008gcn7160,thone2008gcn7161}, and also in radio wave bands
with the Very Large Array \citep[VLA;][]{soderberg2008gcn7178} 
and Westerbork Synthesis Radio Telescope \citep[WSRT;][]{vanderhorst2008gcn7190}.  
As the optical counterpart rapidly brightened, spectroscopic observations revealed broad features
possibly related to an emerging supernova, SN\,2008D, and the source
was first classified as a type Ic SN, but later re-classified to
type Ib based on the emerging presence of helium in the spectra 
\citep[][]{soderberg2008nature,mazzali2008science,malesani2009apj,modjaz2008arxiv}. 
All the SNe associated with GRBs, and the two SNe that have shown mildly relativistic expansion, are type Ic SNe, 
but some type Ib SNe are also expected to produce (at least mildly) relativistic outflows 
\citep[][]{macfadyen1999apj} although this has not been observed so far. 

The nature of the X-ray outburst observed in SN\,2008D has been extensively discussed in the literature. 
It has been claimed that the outburst is a weak XRF caused by a mildly relativistic outflow 
\citep{li2008mnras,mazzali2008science}. 
However, there are counter-claims that we have witnessed the X-ray emission from a supernova shock breakout 
\citep{soderberg2008nature,chevalier2008apj,wang2008aipc}, 
caused by the transition from a radiation dominated to a collisionless shock. 
\citet{soderberg2008nature} have collected 2 months of (high-frequency) radio observations from this source 
and concluded that the outflow is freely expanding at non-relativistic velocities. 
They also argued that the flux of the X-ray outburst in combination with the non-detections at UV/optical wavelengths 
is inconsistent with a (mildly) relativistic outflow. 
On the other hand, the measured variable optical polarization suggests an axisymmetric aspherical expansion 
with variable eccentricity \citep{gorosabel2008arxiv}, as expected in the collapsar model \citep{woosley1993apj}. 
This asphericity has also been inferred from optical spectra of SN\,2008D \citep{modjaz2008arxiv}. 
We note, however, that the optical and radio emission are not unambiguously coming from the same emission region.

In this paper we present the results from our extensive radio follow-up campaigns of SN\,2008D 
with the WSRT and the Giant Metrewave Radio Telescope (GMRT). 
Combined with VLA and CARMA data from \citet{soderberg2008nature}, 
we study these well-sampled light curves up to 17 months after the initial explosion, 
across a broad frequency range, from 325~MHz to 95~GHz. 
We also discuss here the Type Ib SN\,2007uy 
\citep[][]{nakano2008iauc,blondin2008cbet} that went off in the same 
galaxy ten days before the discovery of SN\,2008D. 
Besides radio photometry measurements, we present the results of two epochs 
of global VLBI observations of SN\,2008D and discuss their implications 
for the nature of the source. 
Finally, CO observations with the Arizona Radio Observatory 12-m telescope and lower resolution WSRT measurements 
provide the opportunity to study NGC\,2770, the host galaxy of both SNe, in detail. 
In particular, with two Type~Ib SNe occurring in the same galaxy within 10~days, and three of those SNe within 10~years 
\citep[the third one being SN\,1999eh;][]{hurst1999iauc,jha1999iauc}, 
we discuss the properties of the molecular gas in NGC\,2770 compared to other galaxies, 
which make this galaxy a possible ``SN factory'' \citep[as also discussed in][]{thone2009apj}.

All the measurements of the SNe and their host galaxy are presented in Section~\ref{sec:obs}, 
the modeling and interpretation of the SNe data in Section~\ref{sec:natsne}, 
and of the host galaxy data in Section~\ref{sec:host}. 
In Section~\ref{sec:lofar} we discuss the implications for future observations of such SNe and their host galaxies 
with the Low Frequency Array (LOFAR), the first new generation meter wavelength telescope.
We end with our conclusions in Section~\ref{sec:conclusions}.

% Observations 
\section{Observations}\label{sec:obs}

\subsection{Westerbork Synthesis Radio Telescope}

\begin{figure*}
\begin{center}
\includegraphics[angle=-90,width=\textwidth]{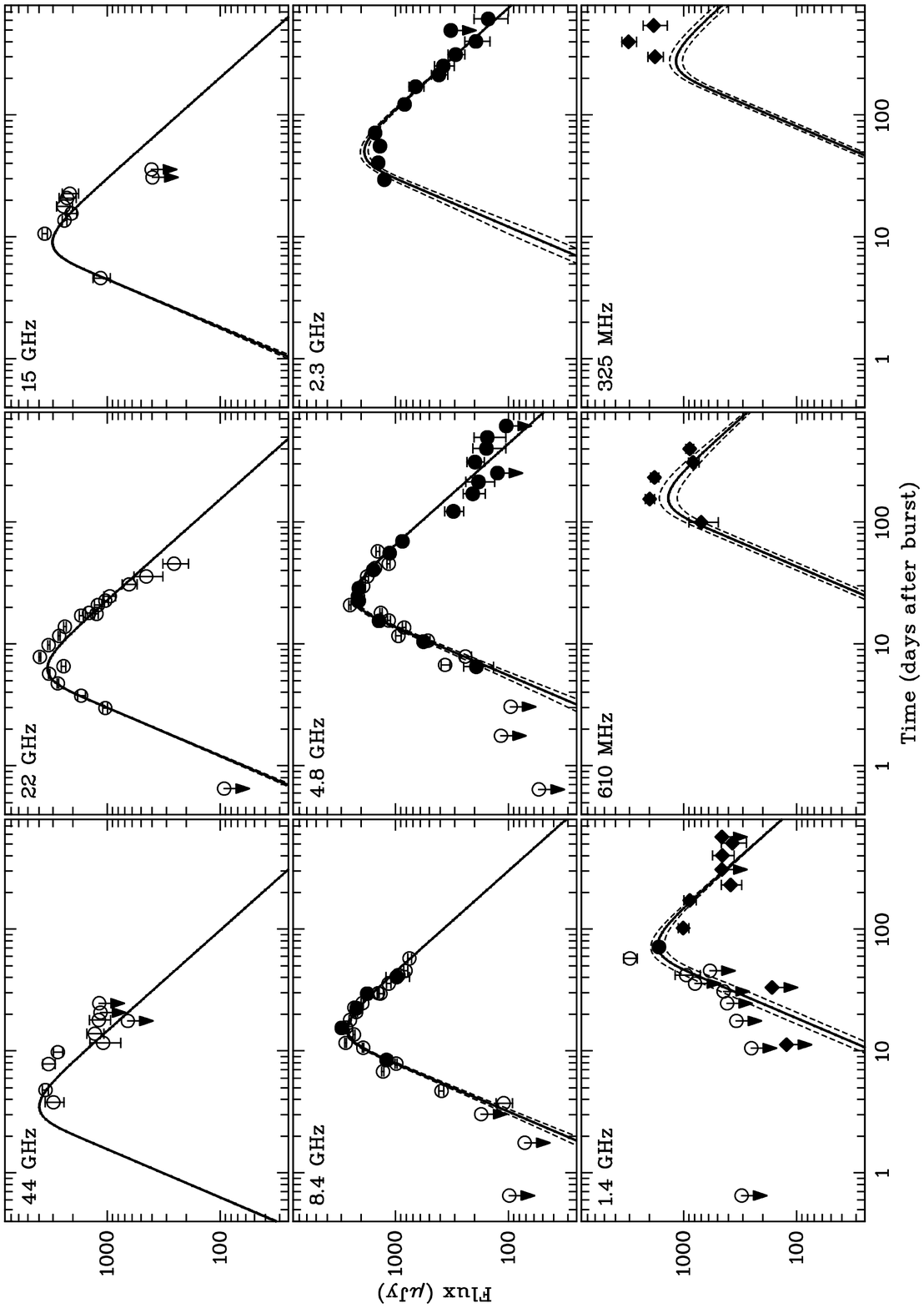}
\caption{Light curves of SN\,2008D from WSRT (solid circles), 
GMRT (solid diamonds), and VLA (open circles); 
VLA fluxes are adopted from \citet{soderberg2008nature}. 
The solid lines represent the best broadband fit to the data, 
while the dashed lines indicate the flux modulation effects of interstellar scintillation. 
The flux values in Table~\ref{table:wsrtgmrtdata} which are lower than three times their flux uncertainty, 
are plotted as $3\sigma$ upper limits.} 
\label{fig:SN2008Dlcs}
\end{center}
\end{figure*}

\begin{figure*}
\begin{center}
\includegraphics[angle=-90,width=\textwidth]{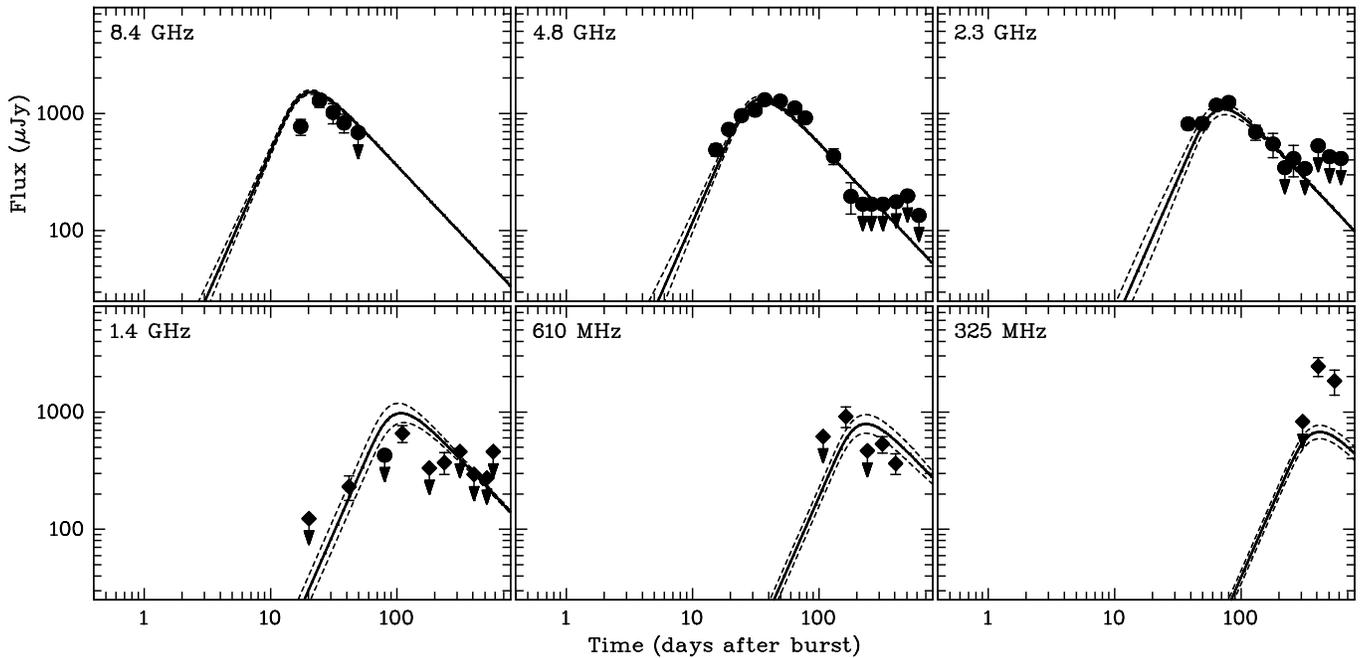}
\caption{Light curves of SN\,2007uy from WSRT (circles) 
and GMRT (diamonds). 
The solid lines represent the best broadband fit to the data, 
while the dashed lines indicate the flux modulation effects of interstellar scintillation. 
The flux values in Table~\ref{table:wsrtgmrtdata} which are lower than three times their flux uncertainty, 
are plotted as $3\sigma$ upper limits.} 
\label{fig:SN2007uylcs}
\end{center}
\end{figure*}

We have performed observations of SN\,2008D and SN\,2007uy with the WSRT 
at 1.4, 2.3, 4.8 and 8.4~GHz. We used the Multi Frequency Front Ends \citep{tan1991} 
in combination with the IVC+DZB back end\footnote[1]{See Section 5.2 at 
http://www.astron.nl/radio-observatory/astronomers/wsrt-guide-observations/wsrt-guide-observations} 
in continuum mode, with a bandwidth of 8x20 MHz at all observing frequencies. 
Gain and phase calibrations were performed with the calibrator 3C~286 for most observations, 
although for a few epochs 3C~48 was used. 
The observations were analyzed using the Multichannel Image Reconstruction Image Analysis and Display 
\citep[MIRIAD;][]{sault1995} software package, except for the WSRT data that were obtained during VLBI observations, 
which were analyzed with the Astronomical Image Processing System \citep[AIPS;][]{wells1985}. 
All the results of our observations, for both SN\,2008D and SN\,2007uy, are detailed in Table~\ref{table:wsrtgmrtdata}; 
the resulting light curves are shown in Figures~\ref{fig:SN2008Dlcs} and \ref{fig:SN2007uylcs}. 

The first observation at 4.8~GHz, at $\sim$6.5 days after the X-ray detection, was reported as a detection of SN\,2008D 
\citep{vanderhorst2008gcn7190}, but the reported flux was significantly higher than the one in Table~\ref{table:wsrtgmrtdata}. 
This discrepancy was caused by a contribution of the host galaxy, which is negligible at 8.4~GHz, 
but significant at 4.8, 2.3 and 1.4~GHz (see Figure~\ref{fig:wsrtmaps}). 
We have now corrected for the diffuse emission of the host galaxy at 4.8 and 2.3~GHz 
by leaving out some of the shortest baselines in generating the radio maps. 
With this technique we lose some sensitivity overall, but we filter out the diffuse emission; 
the longer baselines give us sufficient point-source sensitivity. 
For the 4.8~GHz measurements we had to discard the combinations of telescopes with spacings 
less than $288$~m, while at 2.3~GHz we had to discard spacings less than $576$~m, 
i.e. two times and four times  the $144$~m baseline spacings between the fixed WSRT telescopes, respectively, 
plus the short baselines between the movable telescopes. 
Leaving out these baselines gives radio maps with uniform noise both inside and outside the host galaxy, 
while leaving out more baselines decreases the sensitivity without improving the quality of the images.

We have performed two observations at 1.4~GHz, and we reported the non-detection of a point source 
at the position of SN\,2008D in the first observation \citep{vanderhorst2008gcn7193}, 
including an estimate of the diffuse host galaxy flux at that position of $\sim$1.1~mJy. 
The technique of discarding the shortest baselines did not work at this observing frequency, 
because the length of the baselines that had to be discarded was too large 
for any sensible flux measurement of the two supernovae. 
The fluxes at 1.4~GHz on March 20/21 (Table~\ref{table:wsrtgmrtdata}) 
were obtained by generating difference maps between the March 20/21 observation and the first observation 
at 1.4~GHz on January 18.804 - 19.286. 
This method assumes that the flux of the sources was negligible at the first epoch 
(9.48 and 18.38 days after SN\,2008D and SN\,2007uy, respectively), 
which seems to be a valid assumption when examining the light curves (Figures~\ref{fig:SN2008Dlcs} 
and \ref{fig:SN2007uylcs}). 

\begin{table*}
\begin{center}
\caption{WSRT \& GMRT observations of SN\,2008D and SN\,2007uy}
\label{table:wsrtgmrtdata}
\renewcommand{\arraystretch}{1.1}
\begin{tabular}{|l|l|c|c|c|c|c|c|} 
\tableline
Epoch & Observatory & Duration & Frequency & $\Delta$T$_{\rm{2008D}}$\footnote{$\Delta$T$_{\rm{2008D}}$ is defined from 2008 January 9.564.} & SN\,2008D & $\Delta$T$_{\rm{2007uy}}$\footnote{$\Delta$T$_{\rm{2007uy}}$ is defined from 2007 December 31.669.} & SN2007uy \\
 & & (hours) & (GHz) & (days) & ($\mu$Jy) & (days) & ($\mu$Jy) \\
\tableline
2008 Jan 15.808 - Jan 16.294    & WSRT & 11.7  & 4.8  & 6.48      & 193 $\pm$ 57     & 15.38  & 489 $\pm$ 59 \\
2008 Jan 17.790 - Jan 18.203    & WSRT & 9.9    & 8.4  & 8.43      & 1192 $\pm$ 121 & 17.32  & 772 $\pm$ 121 \\
2008 Jan 19.784 - Jan 20.283    & WSRT & 12.0 & 4.8   & 10.47    & 562 $\pm$ 39     & 19.37  & 728 $\pm$ 54 \\
2008 Jan 20.646 - Jan 20.979    & GMRT & 8.0   & 1.280 & 11.27 & 179 $\pm$ 41\footnote{The source is not significantly detected. We give a formal flux measurement for a point source at the SN\,2008D position.}  & 20.17 & 135 $\pm$ 41\footnote{The source is not significantly detected. We give a formal flux measurement for a point source at the SN\,2007uy position.} \\
2008 Jan 24.770 - Jan 25.270    & WSRT & 5.8   & 4.8   & 15.45    & 1390 $\pm$ 52   & 24.34  & 959 $\pm$ 54 \\
2008 Jan 24.801 - Jan 25.246    & WSRT & 5.2   & 8.4   & 15.46    & 2975 $\pm$ 160 & 24.36  & 1279 $\pm$ 160 \\
2008 Jan 31.751 - Feb 1.251      & WSRT & 5.1   & 4.8   & 22.44    & 2110 $\pm$ 55   & 31.34  & 1065 $\pm$ 69 \\
2008 Jan 31.782 - Feb 1.227      & WSRT & 4.4   & 8.4   & 22.44    & 2199 $\pm$ 197 & 31.34  & 1013 $\pm$ 197 \\
2008 Feb 6.896 - Feb 7.229        & WSRT/VLBI    & 7.7   & 4.8   & 28.50   & 2100 $\pm$ 65   & 37.39  & 1300 $\pm$ 65 \\
2008 Feb 7.732 - Feb 8.231        & WSRT & 5.9   & 2.3   & 29.42    & 1247 $\pm$ 69   & 38.31  & 816 $\pm$ 91 \\
2008 Feb 7.732 - Feb 8.231        & WSRT & 5.3   & 8.4   & 29.42    & 1763 $\pm$ 149 & 38.31  & 828 $\pm$ 149 \\
2008 Feb 11.563 - Feb 11.771   & GMRT & 5.0   & 1.280 & 33.12  & -92 $\pm$ 55     & 42.02 & 231 $\pm$ 55 \\
2008 Feb 18.702 - Feb 19.172   & WSRT & 3.3   & 8.4   & 40.37    & 976 $\pm$ 228   & 49.27   & 660 $\pm$ 228 \\
2008 Feb 18.716 - Feb 19.186   & WSRT & 3.3   & 4.8   & 40.39    & 1567 $\pm$ 61   & 49.28   & 1272 $\pm$ 65 \\
2008 Feb 18.730 - Feb 19.200   & WSRT & 3.3   & 2.3   & 40.40    & 1417 $\pm$ 58   & 49.30   & 820 $\pm$ 98 \\
2008 Mar 4.661 - Mar 5.137        & WSRT & 5.5   & 2.3  & 55.33     & 1363 $\pm$ 61   & 64.22   & 1172 $\pm$ 113 \\
2008 Mar 4.690 - Mar 5.160        & WSRT & 5.5   & 4.8  & 55.36     & 1122 $\pm$ 45   & 64.25   & 1113 $\pm$ 50 \\
2008 Mar 18.792 - Mar 19.146   & WSRT/VLBI    & 8.5   & 4.8   & 69.40  & 862 $\pm$ 66     & 78.30   & 910 $\pm$ 66 \\
2008 Mar 20.617 - Mar 21.117   & WSRT & 5.8   & 2.3   & 71.30    & 1513 $\pm$ 65   & 80.19   & 1233 $\pm$ 90 \\
2008 Mar 20.648 - Mar 21.093   & WSRT & 5.3   & 1.4   & 71.31    & 1644 $\pm$ 139 & 80.20   & 417 $\pm$ 142 \\
2008 Apr 17.372 - Apr 17.774    & GMRT & 9.6   & 0.610 & 99.27   & 696 $\pm$ 205   & 108.17 & 556 $\pm$ 205 \\
2008 Apr 20.454 - Apr 20.773    & GMRT & 7.6   & 1.280 & 102.05 & 1007 $\pm$ 109 & 110.95 & 658 $\pm$ 109 \\
2008 May 10.478 - May 10.864 & WSRT & 4.6   & 2.3  & 122.11   & 826 $\pm$ 59     & 131.00 & 690 $\pm$ 101 \\
2008 May 10.509 - May 10.894 & WSRT & 4.6   & 4.8  & 122.14   & 308 $\pm$ 59     & 131.03 & 431 $\pm$ 65 \\
2008 Jun 12.209 - Jun 12.669   & GMRT & 11.0 & 0.610 & 154.89 & 1986 $\pm$ 182 & 163.79 & 916 $\pm$ 182 \\
2008 Jun 28.344 - Jun 28.814   & WSRT & 5.4   & 2.3  & 171.01   & 659 $\pm$ 98     & 179.91 & 547 $\pm$ 129 \\
2008 Jun 28.373 - Jun 28.842   & WSRT & 5.4   & 4.8  & 171.04   & 207 $\pm$ 46     & 179.94 & 197 $\pm$ 59 \\
2008 Jun 29.228 - Jun 29.452   & GMRT & 5.4   & 1.280 & 171.77 & 880 $\pm$ 111    & 180.67 & 324 $\pm$ 111 \\
2008 Aug 9.230 - Aug 9.700      & WSRT & 5.5   & 2.3   & 212.90  & 411 $\pm$ 67     & 221.80 & 332 $\pm$ 115 \\
2008 Aug 9.258 - Aug 9.728      & WSRT & 5.5   & 4.8   & 212.93  & 185 $\pm$ 53     & 221.83 & 160 $\pm$ 56 \\
2008 Aug 26.184 - Aug 26.333 & GMRT & 3.6   & 1.280 & 229.69 & 385 $\pm$ 78      & 238.59 & 371 $\pm$ 78 \\
2008 Aug 29.155 - Aug 29.378 & GMRT & 5.4   & 0.610 & 232.70 & 1799 $\pm$ 156 & 241.60 & 352 $\pm$ 156 \\
2008 Sep 18.120 - Sep 18.591  & WSRT & 5.5  & 2.3   & 252.79  & 377 $\pm$ 75     & 261.69 & 412 $\pm$ 123 \\
2008 Sep 18.148 - Sep 18.618  & WSRT & 5.5  & 4.8   & 252.82  & 78 $\pm$ 42       & 261.71 & 144 $\pm$ 56 \\
2008 Nov 3.820 - Nov 4.165      & GMRT & 8.3   & 0.325 & 299.46 & 1782 $\pm$ 275 & 308.36 & 786 $\pm$ 275 \\
2008 Nov 11.981 - Nov 12.227 & GMRT & 5.9   & 1.280 & 307.54 & 308 $\pm$ 153   & 316.44 & 208 $\pm$ 153 \\
2008 Nov 12.961 - Nov 13.183 & GMRT & 4.3   & 0.610 & 308.51 & 819 $\pm$ 86   & 317.40 & 532 $\pm$ 86 \\
2008 Nov 15.959 - Nov 16.459 & WSRT & 12.0 & 4.8   & 311.64  & 198 $\pm$ 34    & 320.53 & 155 $\pm$ 56 \\
2008 Nov 16.957 - Nov 17.456 & WSRT & 12.0 & 2.3   & 312.64  & 293 $\pm$ 49    & 321.53 & 266 $\pm$ 113 \\
2009 Feb 9.574 - Feb 9.892      & GMRT & 6.2   & 0.610 & 397.17 & 883 $\pm$ 72   & 406.06 & 365 $\pm$ 72 \\
2009 Feb 10.544 - Feb 10.981  & GMRT & 10.5 & 0.325  & 398.20 & 3046 $\pm$ 444 & 407.10 & 2451 $\pm$ 444 \\
2009 Feb 12.585 - Feb 12.884  & GMRT & 7.2   & 1.280  & 400.17 & 456 $\pm$ 98     & 409.07 & 220 $\pm$ 98 \\
2009 Feb 12.716 - Feb 13.186  & WSRT & 5.5   & 2.3   & 400.39  & 195 $\pm$ 49    & 409.28 & 278 $\pm$ 176 \\
2009 Feb 12.744 - Feb 13.214  & WSRT & 5.5   & 4.8   & 400.41  & 157 $\pm$ 51    & 409.31 & 80 $\pm$ 59 \\
2009 May 16.463 - May 16.933 & WSRT & 5.5   & 2.3   & 493.13  & 208 $\pm$ 108  & 502.03 & 185 $\pm$ 143 \\
2009 May 16.491 - May 16.960 & WSRT & 5.5   & 4.8   & 493.16  & 154 $\pm$ 48    & 502.06 & 48 $\pm$ 66 \\
2009 May 30.708 - May 30.896 & GMRT & 4.5 & 1.280 & 507.24 & 370 $\pm$ 91     & 516.14 & -123 $\pm$ 91 \\
2009 Jul 5.500 - Jul 5.729          & GMRT & 5.5 & 0.325 & 543.05 & 1830 $\pm$ 443 & 551.95 & 1833 $\pm$ 443 \\
2009 Aug 2.604 - Aug 2.854      & GMRT & 6.0 & 1.280 & 571.16 & 457 $\pm$ 153   & 580.06 & -236 $\pm$ 153 \\
2009 Sep 12.138 - Sep 12.637 & WSRT & 12.0 & 4.8   & 611.82  & 47 $\pm$ 35       & 620.72 & 76 $\pm$ 45 \\
2009 Sep 13.139 - Sep 13.634 & WSRT & 11.9 & 2.3   & 612.82  & 151 $\pm$ 50    & 621.72 & 190 $\pm$ 138 \\
\tableline
\end{tabular}
\end{center}
\end{table*}

\subsection{Giant Metrewave Radio Telescope}

We have observed SN\,2008D and SN\,2007uy with the GMRT 
at 1280, 610 and 325~MHz. We have used a bandwidth of 32 MHz for all these observations. 
One of the three possible flux calibrators - 3C48, 3C147, 3C286 - was observed at the beginning 
and end of each observing session for about 15 minutes, 
as a primary flux calibrator to which the flux scale was set.
Radio sources 0741+312, 0842+185 and 0834+555 were used as phase calibrators at 1280, 610 and 325~MHz, respectively. 
Each phase calibrator was observed for about 6 minutes before 
and after an observation of about 30 to 45 minutes on the field centered on SN\,2008D. 
The data were analyzed using AIPS. 

To correct for the diffuse emission due to the host galaxy in the GMRT observations, 
we have removed the shortest baselines, using a similar procedure as for the WSRT measurements. 
The GMRT has a random distribution of 14 out of its 30 antennae within the central square, 
which gives several short baselines. 
We used only those baselines longer than 5 k$\lambda$ at 1280~MHz, 3 k$\lambda$ at 610 MHz, 
and about 2 k$\lambda$ at 325 MHz. 
This ensures discarding the short baselines in the GMRT central square as well as elsewhere in the array. 
All the results of our observations are given in Table~\ref{table:wsrtgmrtdata},
and the light curves are shown in Figures~\ref{fig:SN2008Dlcs} and \ref{fig:SN2007uylcs}.

\subsection{High Resolution Observations with VLBI}

We organized Target of Opportunity global VLBI observations on 2008 February 6 and  
2008 March 18. The primary target was SN\,2008D, 28 and 69 days after the X-ray discovery, 
but short scans on SN\,2007uy were also performed. The participating 
telescopes were Arecibo, Effelsberg, Jodrell Bank (MkII), Hartebeesthoek, Medicina, 
Noto, Onsala, Torun and Westerbork from the European VLBI Network (EVN), and Hancock and St.~Croix from the VLBA. 
The  8-hour observations (of which Arecibo could track the source only
for about 1.5 hours) were carried out at 5~GHz at 1024 Mbps using 2-bit sampling.
The VLBA stations recorded at a rate of 512 Mbps with 1-bit sampling, to obtain the
same observing bandwidth. The synthesis array data from the WSRT were recorded parallel to
the VLBI observations. The target was phase-referenced to the nearby calibrators
J0911+3349 (C1) and J0919+3324 (C2), at an angular distance of 0.8 and 2~degrees from the target (T), respectively. 
The phase reference cycle pattern was T-C1-T-C1-C2 etc. with corresponding 3:30-1:30-3:30-1:30-2:30 minute scans; 
the last scan included a 1 minute gap for slewing and system temperature measurements at the EVN telescopes. 
The total on source time on SN\,2008D was 200 minutes and 210 minutes at the two epochs, respectively. 
As a comparison source, we included scans on SN\,2007uy phase-referenced in a similar fashion. 
The total on-source time on SN\,2007uy was 38 minutes at the first epoch, and 49 minutes at the second epoch.

The data reduction was carried out in AIPS using standard techniques \citep[e.g.][]{diamond1995}, 
and the calibrated data were exported to Difmap \citep[][]{shepherd1994baas}. 
Because J0911+3349 had a resolved structure, its structural phase was 
removed in the process of phase-referencing. First we phase-referenced
J0911+3349 to J0919+3324, and made a map of these nearby calibrators in 
Difmap. We carried out the structural phase correction in two different ways. 
The fringe-fit solutions from J0919+3324 were interpolated to all sources.
The J0911+3349 data were additionally phase self-calibrated and 
these solutions, too, were interpolated to the target. In the other 
method we repeated fringe-fitting from scratch, using now J0911+3349 as 
the primary reference source and its map was used to correct for the 
structural phase. These two methods should ideally give 
identical or very similar results. The advantage of the first method is that
the compact source J0919+3324 is expected to produce higher 
signal-to-noise delay and delay-rate solutions, but the phase-reference
cycle time in this case is quite long, i.e. 11 minutes. However, to determine 
the target source position independent of possible remaining structural phase and 
positional errors in the nearby calibrator J0911+3349, we directly phase-referenced 
the target to J0919+3324. The coordinates of this source were taken from the 
VLBA Calibrator Survey\footnote[2]{http:/$\!$/www.vlba.nrao.edu/astro/calib/index.shtml}: 
RA=$09^{\rm h}19^{\rm m}08^{\rm s}.787122$, Dec=+33$^{\circ}$24\arcmin41\arcsec.94287 (J2000).

The global VLBI images of SN\,2008D appeared unresolved at both epochs. 
The observed peak brightnesses were 2.0~mJy and 0.9~mJy on February 6 
and March 18, respectively. These values are consistent with the WSRT 
total flux density measurements of 2.09$\pm$0.06 mJy and 0.86$\pm$0.06 
mJy, taken during the VLBI observations (see Table~\ref{table:wsrtgmrtdata}). 
The off-source image noise was 20-25 $\mu$Jy/beam using natural weighting, 
but it was higher near the target. 

We carried out model-fitting of the $uv$-data in Difmap to give constraints
on the apparent angular size. Both point and circular Gaussian components
were fitted to the data. At the first epoch, we obtained angular diameter sizes
of 0.36 and 0.40~milli-arcsecond (mas) using the two different ways of processing. The 
second epoch data produced different results. Using the first method the
circular Gaussian component collapsed to zero radius; in the other case
we obtained a size of 0.5 mas, although the fit was poorer.
The reason for this difference may be the poor phase stability at the 
second epoch due to bad weather conditions. J0911+3349, after
phase-referencing to the more compact calibrator, showed phase 
instabilities on short timescales. Model-fitting of the sparser SN\,2007uy 
data at the first epoch resulted in a similar, although somewhat smaller size
of 0.28 mas and 0.12 mas using the two calibration methods. At the second epoch 
the size of the fitted circular Gaussian component collapsed to zero radius
in both cases.

To better understand the significance of the model-fitting results for SN\,2008D, we performed Monte Carlo 
simulations. Using simulated data with $uv$-coverage and telescope sensitivites identical to the first
epoch observations, 2~mJy circular Gaussian sources with a size of 0.33~mas and 0.03~mas (practically a point 
source) were added to the simulated data in 400 trials each. 
In the first case the recovered source sizes were typically 0.3~mas and in all cases less than 0.5~mas. 
In the second case with the point source model, the recovered size is less than 0.25~mas. 
These simulations indicate that the data are marginally consistent with a resolved
source of 0.36--0.40 mas with an error of $\pm$0.1 mas. 
We note, however, that in this analysis we simulated only the effect of thermal noise, 
and the error may be somewhat higher due to adverse weather 
and other systematic errors that may have affected phase-referencing. 
For the second epoch we simulated a 0.9 mJy source with a size of 0.35 mas to see if it could be consistent with
the previous epoch and the measured value. 
The results were very similar to the first epoch, with recovered sizes less than 0.5~mas. 
We conclude that at the second epoch the sizes obtained via the two different processing methods 
are consistent with a source size of a few tenths of mas. 
Considering the poor weather conditions described above, and the fact that in all cases the 
point models provided equally good fits, we consider the 0.4~mas and 0.5~mas values as upper limits for the 
angular diameter size of SN\,2008D on February 6 and March 18, respectively. 
These upper limits are comparable to the ones obtained by \citet[][]{bietenholz2009apj} at 8.4~GHz on February 8, 
and at 5.0 and 8.4 GHz on May 21, assuming a spherical shell model for the supernova ejecta.

The position for SN\,2008D is RA=$09^{\rm h}09^{\rm m}30^{\rm s}.646311$, Dec=+33$^{\circ}$08\arcmin20\arcsec.123445 (J2000) 
at the first epoch, and RA=$09^{\rm h}09^{\rm m}30^{\rm s}.646318$, Dec=+33$^{\circ}$08\arcmin20\arcsec.123473 (J2000) at the second epoch. 
The difference between these two epochs in X (RA cos(Dec)) is 0.088 mas and in Y (Dec) is 0.028 mas,
well within the expected astrometric accuracy of $0.10-0.15$ mas \citep[][]{pradel2006aa}, consistent with no apparent
proper motion of the source. We note, however, that these positions seem to be systematically different
by almost 1 mas from the declination values obtained by \citet{bietenholz2009apj}. There are three
contributing factors that may explain this discrepancy: (i) there was a slight difference in the coordinates
of J0919+3324 used ($\sim81~\mu$as); (ii) the observing frequencies were different and a frequency-dependent core-shift of a 
few tenths of mas may be present in the reference source; (iii) different models were used for calculating the 
tropospheric delay at the two correlators. We varied the tropospheric zenith delay of all telescopes by 
$\pm10$~cm before fringe-fitting to investigate the effect of the poorly modeled troposphere. We found
that the resulting target positions between these two extremes differed by about 0.5 mas. We thus conclude 
that the level of systematic errors (besides the position of the reference source) may be a few hundredths of mas. 

For SN\,2007uy we did not perform detailed simulations on the source size because of the worse $uv$-coverage and sensitivity, 
and the source being fainter. Hence the Monte Carlo simulations would show a broader source size distribution 
than in the case of SN\,2008D, and since the source size is somewhat smaller (0.28~mas), we conclude 
that SN\,2007uy is consistent with a point source with an upper limit on the angular diameter size of 0.28~mas in the first epoch. 
For the second epoch, a reliable upper limit on the angular diameter size could not be obtained. 
In the case of SN\,2007uy there is also no significant proper motion, with a much larger uncertainty in the positions than SN\,2008D, 
again due to the worse $uv$-coverage, sensitivity and source faintness. The most accurate source position is obtained at the first epoch: 
RA=$09^{\rm h}09^{\rm m}35^{\rm s}.300206$, Dec=+33$^{\circ}$07\arcmin09\arcsec.007559 (J2000), 
with an uncertainty of several hundreds of $\mu$as.

\subsection{Arizona Radio Observatory 12-m Telescope}

\begin{figure*}
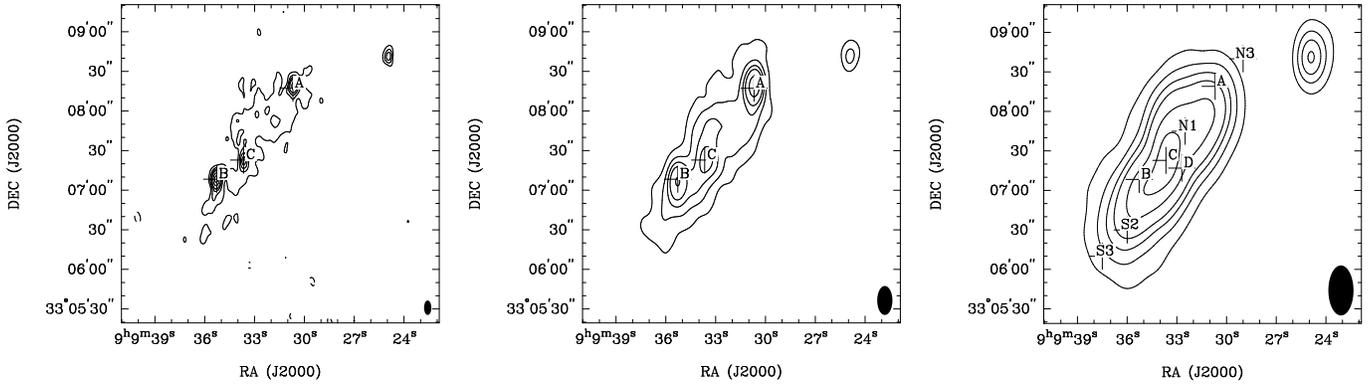

\begin{center}
\includegraphics[angle=-90,width=0.32\textwidth]{figure3a.ps}
\hspace{0.01\textwidth}
\includegraphics[angle=-90,width=0.32\textwidth]{figure3b.ps}
\hspace{0.01\textwidth}
\includegraphics[angle=-90,width=0.32\textwidth]{figure3c.ps}
\caption{WSRT maps of the NGC\,2770 field at 4.8~GHz (left), 2.3~GHz (middle), 
and 1.4~GHz (right), with the beam size plotted in the lower right corner of each panel. 
Indicated are SN\,2008D (A), SN\,2007uy (B), the core of NGC\,2770 (C), and SN\,1999eh (D). 
On the 1.4~GHz map the ARO observation positions are also shown (see Table~\ref{table:12data} for the coordinates). 
Radio brightness contours are given at the -3, 3, 6, 9, 12, 18, 24, 30 sigma levels. 
We note that the compact source in the right hand corner of each panel has a spectral index of $-1.2$, 
and is not significantly variable over our $\sim600$~days of observations.} 
\label{fig:wsrtmaps}
\end{center}
\end{figure*}

We obtained CO $J=1\to 0$ data  of NGC\,2770 at an observing frequency of 115.2712~GHz on 2008 February 3-4, 
using the ARO 12-m telescope, located at Kitt Peak National Observatory. 
We used the standard combination of beam and position switching (BSP) with the 3~mm~HI receiver 
(dual polarization) connected to the 2~MHz filterbanks. 
We observed a total of eight positions, which are given in Table~\ref{table:12data} and shown 
in the right panel of Figure~\ref{fig:wsrtmaps} on top of the 1.4~GHz map of the galaxy. 
Each supernova position was observed, as were the center, the positions designated S2, S3, N1 and N3, 
where the letter indicates south or north, and the number of $55\arcsec$ beams south or north of the center, along the major axis of the galaxy. 
Supernova 2007uy occurred very close to what would otherwise be designated the S1 position, and SN 2008D at approximately the N2 position. 
Supernova 1999eh was about half a beam from the center, approximately along the minor axis. 

\begin{figure}
\begin{center}
\includegraphics[width=\columnwidth]{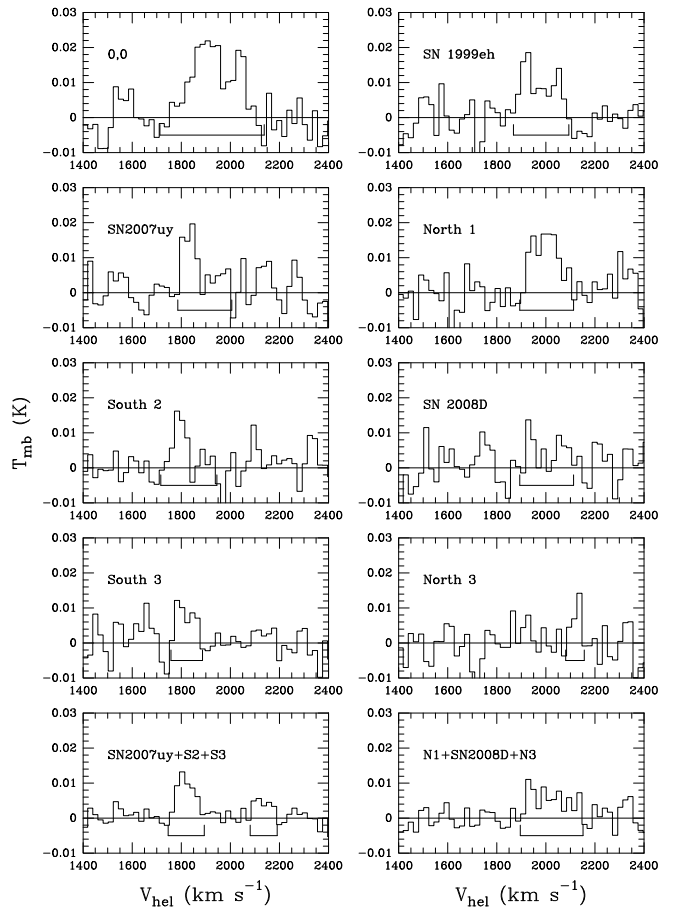}
\caption{CO $J=1\to 0$ added and smoothed spectra, with positions indicated. The solid lines below the emission indicate the line windows used to determine the integrated intensities.} 
\label{fig:n2770-co}
\end{center}
\end{figure}

The final spectra are displayed in Figure~\ref{fig:n2770-co}, where they have been Hanning smoothed twice 
to a resolution of 20.8 km s$^{-1}$ per channel, and scaled to the main beam temperature scale 
using a main beam efficiency $\eta_{\rm MB}=0.60$ in order to calculate the total mass of H$_2$ and He. 
We assume a standard conversion factor of $N({\rm H}_2$) = $I_{\rm MB}\,2.3 \times 10^{20}$ mol cm$^{-2}$ K km s$^{-1}$ 
\citep[see e.g.][]{sage2007apj}, with $I_{\rm MB}$ the main beam integrated intensity. 
The {\it observed} integrated line intensity in units of antenna temperature $T_A^*$ is given in column 5 of Table~\ref{table:12data}, 
while column 8 contains the molecular Hydrogen (H$_2$) mass derived from the scaled main beam integrated intensities 
(no mass is given for the SN1999eh position because the beam overlaps strongly with the center position and it is therefore not physically meaningful). 

\begin{table*}
\begin{center}
\caption{ARO 12m observations of CO emission from NGC\,2770}
\label{table:12data}
\renewcommand{\arraystretch}{1.1}
\begin{tabular}{|l|l|c|c|c|c|c|c|} 
\tableline
R.A. (J2000) & Dec (J2000) & Position & Time & $I_{\rm CO}$ & $\sigma$ & window & $M$(H$_2$) \\
(hr:min:sec) & (deg:min:sec) & & (min) & (K km s$^{-1}$) & (K km s$^{-1}$) & (km s$^{-1}$) & ($10^8$ M$_\odot$) \\
\tableline
09:09:33.7    &  33:07:25.0 & 0  & 36 & 2.60  &  0.34  & 1713 2138 & 12 \\
09:09:32.7 & 33:07:17.0    & SN 1999eh & 36  & 1.27 & 0.22  & 1868 2094 & ... \\
09:09:34.4   & 33:07:09.9 & SN 2007uy & 36  & 1.05 & 0.24 & 1786 2006 & 5.8 \\
 09:09:32.5 & 33:07:45.0 & N1   & 36  & 1.37  & 0.23   & 1894 2113 & 7.5 \\
09:09:36.0 &  33:06:30.0 & S2   & 36  & 0.56   & 0.20 & 1717 1945  &  3.5\\
09:09:30.65 & 33:08:20.3 & SN 2008D & 36   &  0.62   &  0.22 &  1913 2109  & 3.3 \\
09:09:37.5 & 33:06:10.0 & S3 & 48 & 0.56   & 0.16 & 1758 1887 & 3.1 \\
09:09:29.0 & 33:08:40.0 & N3 & 60 & 0.32   & 0.11 & 2081 2157 &  1.8 \\
  &   &   S1+S2+S3 & &  0.63  & 0.08 & 1747 1894  &  3.5 \\
  &   &   S1+S2+S3 & &  0.28  & 0.07  & 2081 2191 &  1.5 \\
  &   &   N1+N2+N3 & &  0.85  & 0.14 & 1896 2152  &  4.8 \\
\tableline
\end{tabular}
\end{center}
\end{table*}

% Nature of SNe 2008D and 2007uy
\section{Nature of SNe 2008D and 2007uy}\label{sec:natsne}

SN\,2007uy and SN\,2008D have both been detected from radio to X-ray frequencies. 
The discovery of an X-ray burst associated with SN\,2008D has led to much more detailed studies of this source compared to SN\,2007uy, 
especially because of the possible presence of a (initially) relativistic jet causing the X-ray emission 
(see Section \ref{section:intro} for a detailed discussion). 
Broadband observations have been carried out by several groups, starting immediately after the X-ray outburst. 
Here we will discuss our modeling and results of our extensive radio follow-up campaign.

\subsection{Modeling of SN Light Curves}\label{subsec:model}

The radio light curves of both SNe, shown in Figures \ref{fig:SN2008Dlcs} and \ref{fig:SN2007uylcs}, 
display the typical general behavior of the peak of a synchrotron emission spectrum moving through the observing bands 
from high to low frequencies, as expected for both radio SNe and GRB afterglows. 
The light curves are determined by the shape of the spectrum and the evolution of the peak, 
which in turn are determined by the nature and evolution of the ejecta. 
In GRB afterglows the spectral peak either corresponds to the minimum energy of the synchrotron emitting electrons 
(with a corresponding peak frequency $\nu_{\rm{m}}$), 
or to a turnover caused by synchrotron self-absorption \citep[with self-absorption frequency $\nu_{\rm{a}}$; e.g.][]{sari1998apj,wijers1999apj}. 
In both cases there is a high-energy optically thin spectral index $\beta$, 
but a fixed low-energy spectral index with a value of $1/3$~($\nu_{\rm{a}}<\nu<\nu_{\rm{m}}$), 
$2$~($\nu<\nu_{\rm{a,m}}$) or $2.5$~($\nu_{\rm{m}}<\nu<\nu_{\rm{a}}$). 
Radio SNe can also be described with synchrotron self-absorption \citep{chevalier1998apj}, 
or with free-free absorption from the ambient medium, resulting in an exponential cutoff below the peak \citep{weiler1986apj}. 

\begin{figure}
\begin{center}
\includegraphics[angle=-90,width=\columnwidth]{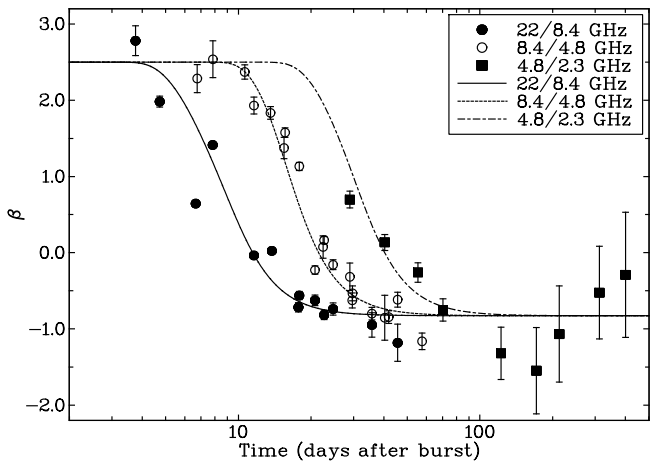}
\caption{Spectral indices of SN\,2008D based on the observed flux ratios between 22 and 8.4~GHz (solid circles), 
8.4 and 4.8~GHz (open circles), 4.8 and 2.3~GHz (solid squares). 
The lines show the evolution of the spectral indices based on the model fits shown in Figure~\ref{fig:SN2008Dlcs}.} 
\label{fig:SN2008Dspecind}
\end{center}
\end{figure}

Figure~\ref{fig:SN2008Dspecind} displays the evolution of the spectral indices for SN\,2008D, 
based on the flux ratios between 22 and 8.4~GHz \citep[data from][]{soderberg2008nature}, 
8.4 and 4.8~GHz \citep[data from this paper and][]{soderberg2008nature}, and 4.8 and 2.3~GHz (data from this paper). 
Figure~\ref{fig:SN2008Dspecind} shows a smooth transition from an optically thick spectral index of $\beta\sim 2.5$ 
to an optically thin index $\beta\sim -0.83$. 
The latter value has been determined in the broadband radio light curve modeling described below. 
Given these spectral indices, the evolving spectrum of SN\,2008D can be well described by synchrotron emission with synchrotron self-absorption. 
Our early-time coverage of SN\,2007uy is too sparse to determine the optically thick spectral index in a similar way as for SN\,2008D, 
but the optically thin part is consistent with $\beta\sim -0.8$. 
Therefore, in the following we also adopt a synchrotron self-absorbed spectrum for SN\,2007uy.

Now that we have established the shape of the spectrum, we model the light curves at all observing frequencies simultaneously 
by letting the synchrotron self-absorped spectrum evolve in time. 
In our modeling the synchrotron self-absorption frequency $\nu_{\rm{a}}$ 
and the peak flux $F_{\nu,{\rm{a}}}$ decay as power laws in time, which provides a good broadband fit to the radio data of both SNe. 
The resulting model light curves for SN\,2008D are shown as solid lines in Figure~\ref{fig:SN2008Dlcs}. 
From our modeling, we find that $\nu_{\rm{a}}\propto t^{-1.1}$ and $F_{\nu,{\rm{a}}}\propto t^{-0.28}$, 
and an optically thin spectral index $\beta$ of $-0.83$. 
The latter implies an electron energy distribution index $p=-(2\beta+1)\sim 2.7$. 
This value for $p$ is comparable to those of Ib/c SNe \citep[e.g.][]{chevalier2006apj} 
and of GRB afterglows \citep[e.g.][]{panaitescu2002apj}. 

The expected temporal behavior of $\nu_{\rm{a}}$ and $F_{\nu,{\rm{a}}}$ depends on the value of $p$ 
and the evolution of the ejecta, which is different for young radio SNe and for decelerating GRB shocks. 
In the case of Ib/c SNe the ejecta are freely expanding, with the radius $R$ of the shock, 
which is formed by the interaction of the ejecta with the ambient medium, roughly proportional to $t$. 
The external shock in GRBs is decelerating when it is producing the radio emission, 
giving a different behavior of $R$ as a function of $t$, depending on whether the shock velocity is relativistic or non-relativistic 
and on the density structure of the surrounding medium (either homogeneous or structured by a massive stellar wind). 
When we assume that the radio emission in SN\,2008D comes from a decelerating shock, 
the observed temporal evolution of $\nu_{\rm{a}}$ and $F_{\nu,{\rm{a}}}$ and the value of $p$ 
are not consistent with the expected relations between those parameters. 
The evolution of $\nu_{\rm{a}}$ would be consistent with the observed $p$ value 
if the shock was ploughing through a stellar wind medium, 
but the inferred temporal indices for $F_{\nu,{\rm{a}}}$ are too steep 
($-0.9$ and $-1.3$ respectively) compared to the observed one ($-0.28$). 
Thus, the spectrum and light curves of SN\,2008D are inconsistent with emission from a decelerating shock, 
but they can be explained in the context of SN ejecta which are freely expanding into a stellar wind 
\citep[consistent with the findings of][]{soderberg2008nature}. 
For SN\,2007uy we find similar temporal indices as for SN\,2008D, $\nu_{\rm{a}}\propto t^{-1.1}$ and $F_{\nu,{\rm{a}}}\propto t^{-0.27}$, 
which are also typical of freely expanding SN ejecta and inconsistent with a decelerating shock (Figure~\ref{fig:SN2007uylcs}). 

Based on our best fit light curves (Figures~\ref{fig:SN2008Dlcs} and \ref{fig:SN2007uylcs}) we have determined 
the kinetic energy $E$, radius $R$, post-shock magnetic field strength $B$, and density $\rho$ of the radio ejecta for both SNe. 
We have adopted the prescription of \citet[][]{soderberg2005apj}, assuming equipartition between the magnetic field 
and kinetic energy of the electrons ($\epsilon_{\rm{e}}=\epsilon_{\rm{B}}$=0.1). 
We also assume that the characteristic synchrotron frequency $\nu_{\rm{m}}$, corresponding to the minimum energy 
in the electron energy distribution, is 1~GHz at 10 days after the SNe explosions, 
but we note that the physical parameters depend very weakly on $\nu_{\rm{m}}$. 
Since the surrounding medium is structured by a massive stellar wind, the mass density is given by $\rho=A\cdot R^{-2}$. 
For SN\,2008D we obtain $E=6.7\times10^{47}~t_{10}^{\,0.5}$~erg, $R=4.9\times10^{15}~t_{10}^{\,0.94}$~cm, 
$A_*=0.18$ and $B=3.4~t_{10}^{\,-1.1}$~G, 
while for SN\,2007uy we find $E=1.4\times10^{47}~t_{10}^{\,0.7}$~erg, $R=2.6\times10^{15}~t_{10}^{\,0.96}$~cm, 
$A_*=0.076$ and $B=4.0~t_{10}^{\,-1.1}$~G, with $t_{10}=t/(10~\rm{days})$ and $A_*$ is $A$ in units of $5\times10^{11}$~g~cm$^{-1}$. 
The physical parameters we obtained for these SNe are within the ranges of the parameter values that have been found for other 
core-collapse SNe \citep[e.g.][]{fransson1998apj,berger2002apj,soderberg2005apj,soderberg2006apj2}. 
In the case of SN\,2008D they are comparable with the results from \citet{soderberg2008nature}.

\subsection{Constraints on an Off-Axis Jet}

Our long-term observations of SN\,2008D and SN\,2007uy can also 
be used to constrain the possible existence of an off-axis (mildly) 
relativistic jet, which becomes visible only at late times.  To this end, we 
broadly follow the method of \citet[][]{granot2003apj}, and generalize 
the results of \citet[][]{nakar2002apj} from a uniform density to a
density that drops as $\rho=A\cdot R^{-2}$. For a random orientation of a
bipolar jet, the typical viewing angle to the jet closest to us 
is fairly large, $\theta_{\rm obs} \sim 1$. Thus, for simplicity, we assume 
such a large viewing angle, for which the jet becomes visible near the
non-relativistic transition time, $t_{\rm NR}$. Then, we 
attribute the excess at 325$\;$MHz around $t = 400\;$days, when the flux
is $F_\nu \approx 3\;$mJy, as the flux from such a jet around the time
when it becomes visible ($\sim t_{\rm NR}$ of that jet).  This serves
two purposes: (i) we might possibly be starting to see a contribution
from such an off-axis jet, and (ii) even if this is not the correct
explanation for this flux excess (which might very well be the case), 
then this exercise would still give us a handle (or limits) on 
the properties of an off-axis jet that could hide beneath the
observed flux level.

If we adopt the value of $A_* \sim 0.7$, which was inferred by
\citet[][]{soderberg2008nature}, then such a relatively large external
density requires a large jet energy (of a few $10^{51}\;$erg) in order
to have a late ($\sim 400\;$days) peak time. This, in turn, would
require very low values for the shock microphysical parameter
(e.g. $\epsilon_B \sim 10^{-4}$ and $\epsilon_e \sim 0.02$) in order
not to exceed the observed flux level.

Alternatively, if the density along the rotational axis is
significantly lower than near the equator, then $A_* \sim 0.03$ along
the axis would allow a more reasonable solution. As an illustrating
example, one possible solution for the remaining model parameters is
$\epsilon _e \sim 0.1$, $\epsilon_B \sim 0.002$ and $E \sim 1.7\times
10^{50}\;$erg (the value of $p$ is hardly constrained by the peak time
and flux, and could be estimated from the observed spectral slope).

\subsection{Constraints from VLBI Observations}

We have constrained the size of the radio ejecta from SN\,2008D at early epochs. 
The source was unresolved in our images, with upper limits 
on the angular diameter size of 0.4~mas at 28~days and 0.5~mas at 69~days. 
These angular sizes result in upper limits on the average apparent isotropic expansion velocity of $1.1c$ for the first 28 days 
and $0.57c$ for the first 69 days after the stellar explosion. 
From the upper limit on the proper motion of the radio source, we derive an average expansion velocity between the two epochs of $0.38c$. 
For SN\,2007uy there is only a reliable upper limit on the angular diameter size at 37~days of 0.3~mas, 
resulting in an upper limit on the average expansion velocity of $0.64c$. 

These upper limits for both SNe show that if there were a jet, it was moderately relativistic. 
This result is consistent with the analysis of the radio light curves of this source, 
and also in agreement  with conclusions derived from VLBA and HSA observations 
by \citet{soderberg2008nature} and \citet{bietenholz2009apj}.

\subsection{Interstellar Scintillation Effects}

The light curves in Figures~\ref{fig:SN2008Dlcs} and \ref{fig:SN2007uylcs} show that some data, 
in particular at the lower radio frequencies, deviate significantly from the best-fit light curves. 
We have investigated whether this could be due to interstellar scintillation (ISS), 
i.e. the effect of the interstellar medium modulating the radio fluxes of sources that have angular sizes which are smaller than the typical scintillation scales. 
There are three types of ISS, i.e. weak, refractive and diffractive. 
For SN\,2008D and SN\,2007uy the effects of weak scintillation are very small, at most a few percent, 
while diffractive scintillation plays an even less significant role. 
The effects of refractive ISS are much larger, ranging up to several tens of percent, depending on the observing frequency and time. 
Details of the ISS calculations are given in Appendix \ref{sec:appendix}. 
The ISS effects are shown in Figures~\ref{fig:SN2008Dlcs} and \ref{fig:SN2007uylcs} as dashed lines, 
showing the maximum and minimum modulations of our best fit model fluxes. 
It is evident in the Figures that ISS can explain most of the scatter in the observed fluxes of the two SNe, 
except for the 325~MHz measurements at $\sim 400$ days. 
From the current data-set it is not clear if the latter deviations are intrinsic or caused by systematic observational effects. 
The temporal behavior of the modeled light curves is mainly determined by the better sampled higher radio frequencies, 
and deviating behavior at later times and lower frequencies is certainly possible. 
However, the fact that both SNe display the same high flux could be an indication of systematics, 
but future 325~MHz observations with the GMRT and at even lower frequencies with LOFAR (see Section~\ref{sec:lofar}) 
will be able to solve this issue.

% The Host Galaxy NGC\,2770
\section{The Host Galaxy NGC\,2770}\label{sec:host}

SN progenitors are young stars on galactic timescales, 
so it is relevant to consider the molecular component of the galactic gas where these stars form. 
Although the linear resolution of the single dish observations is $\sim$8.6~kpc 
(and therefore much larger than single giant molecular clouds), 
general statements about the star forming environment are still possible. 

There are several striking features in Figure~\ref{fig:n2770-co}. 
The gas distribution is, unlike in most spiral galaxies, asymmetric. 
The center position has an asymmetric shape around the systemic velocity of 1947~km/s. 
There is more gas at the N1 position than at SN\,2007uy (=S1), and more at S2 than at SN\,2008D (=N2), 
while the SNe have occurred far too recently to affect molecular clouds on the scale sampled by the 12m beam. 
To explore this asymmetry further, we co-added the north and south positions separately; 
the resulting spectra are the bottom left and right panels in Figure~\ref{fig:n2770-co}. 
The most interesting feature appears in the lower left of the Figure (total of the south positions). 
The gas following the galaxy's CO rotation curve (centered around $\sim$1820~km/s) is obvious, 
and typical of normal spiral galaxies \citep[see e.g.][]{sage1993aa}. 
A second feature, centered around $\sim$2125~km/s, is also visible. 
The statistical significance of the feature is $4\sigma$. 
The physical significance, if one accepts the $4\sigma$ detection, is that this indicates a counter-rotating component of gas. 
No such counter-rotating feature is seen in the summed north positions (lower right panel of Figure~\ref{fig:n2770-co}).

\subsection{A comparison of CO and HI}

To explore the significance of particular features in the CO results more thoroughly, 
we compared the data to the HI line observed by \citet{matthews2001aa} using the Nan\c{c}ay telescope in France, 
and the WSRT HI map from \citet{rhee1996aas}. 
The HI line from \citet[][their Figure 1]{matthews2001aa} shows the same asymmetry as the CO center position above. 
Even more striking is the map of \citet{rhee1996aas}, which clearly shows the HI rotating in the {\it opposite} sense of the CO, 
with high velocities south of the center. 
At the 4$\sigma$ level (second positive contour) there is a counter-rotating HI component in the south, 
at low velocities (centered around $\sim1850$~km/s). 
This counter-rotating HI is therefore co-rotating with the CO, although the direction of rotation of the stars has not been reported, 
but the usual assumption is that the HI and stars co-rotate. 
There is only one other galaxy for which counter-rotating gas components have been reported, namely NGC\,4546 \citep[][]{sage1994aj}. 

The only plausible source of counter-rotating gas is a collision or merger. 
NGC\,2770 does have a dwarf companion, NGC\,2770B, and based upon recent work on the local group of galaxies 
\citep[e.g.][]{mcconnachie2009nature}, it is clear that collisions with dwarf galaxies are ongoing into the present epoch. 
We accordingly conclude that NGC\,2770 has absorbed a dwarf galaxy at some point in the last $\sim10^8$~yr or so. 
The resulting cloud-cloud collisions have induced star formation, and we are now seeing the resulting SNe. 
Although the SNe have occurred in different parts of NGC\,2770, studies of other galaxies have shown that collisions or mergers 
can affect the star formation in the entire galaxy \citep[see e.g.][]{bureau2006mnras}.

Using the inclination and absorption corrected UBV colors, it is possible to estimate (rather crudely, and with a lot of assumptions) 
how long ago a burst of star formation began \citep[][]{larson1978apj}. 
The UBV colors of NGC\,2770, compared to Figure~2b (appropriate for a blue galaxy) of \citet[][]{larson1978apj}, 
lie in the position of a burst that occurred $\sim10^8$~yr ago that provided about 10 percent of the current number of stars, 
in a burst lasting $2\times10^7$~yr. 
If the SNe indeed result from a merger-induced burst of star formation, we are observing the tail of that burst, 
which spreads out over time and space. Individual clouds trigger localized bursts, 
but the cloud collisions will continue for $\sim10^9$~yr until the system has equilibrated.

Although NGC\,2770 has a rather high gas mass and HI surface density compared to other nearby spiral galaxies, 
indicating that there is a lot of material present to produce stars, 
the star formation rate does not seem to be exceptionally high, 
and the same is true for the local star formation rate at the sites of the SNe \citep[][]{thone2009apj}. 
However, the star formation rate of NGC\,2770B appears to be large; 
in fact, the specific star formation rate of NGC\,2770B is very high and it is one of the most metal poor galaxies ever detected, 
indicating a very young stellar population \citep[][]{thone2009apj}. 
Although the specific star formation rate of NGC\,2770B is larger than for NGC\,2770, 
there have not been any SNe detected in NGC\,2770B, which can be explained by the fact that 
we are just taking an instantaneous snapshot compared to the $10^7-10^8$~yr timescale 
at which SNe trail star formation.

Counter-rotating gas clouds will collide much more often than co-rotating gas clouds, leading to collision-induced star formation. 
Whether the SN activity in NGC\,2770 is striking or not, 
with 2 core-collapse SNe within 10 days and 3 of them within 9 years, 
it is certainly plausible that we are seeing the result of a collision-induced burst of star formation.

% Implications for LOFAR
\section{Future Low Frequency Observations }\label{sec:lofar}

The peak of the spectra of both SN\,2007uy and SN\,2008D has now moved below the lowest frequency 
at which these two sources have been observed so far, namely at 325~MHz with the GMRT. 
A new generation of meter wavelength radio telescopes will now be required to continue to observe these SNe and also many other radio SNe. 
The Low Frequency Array (LOFAR), a major new multi-element, 
interferometric, imaging telescope designed for the 30-240~MHz range \citep[see e.g.][]{rottgering2006arxiv}, 
will have unprecedented sensitivity and resolution at meter wavelengths. 
It is currently being built and will be fully functioning at the end of 2010.

\begin{figure}
\begin{center}
\includegraphics[angle=0,width=\columnwidth]{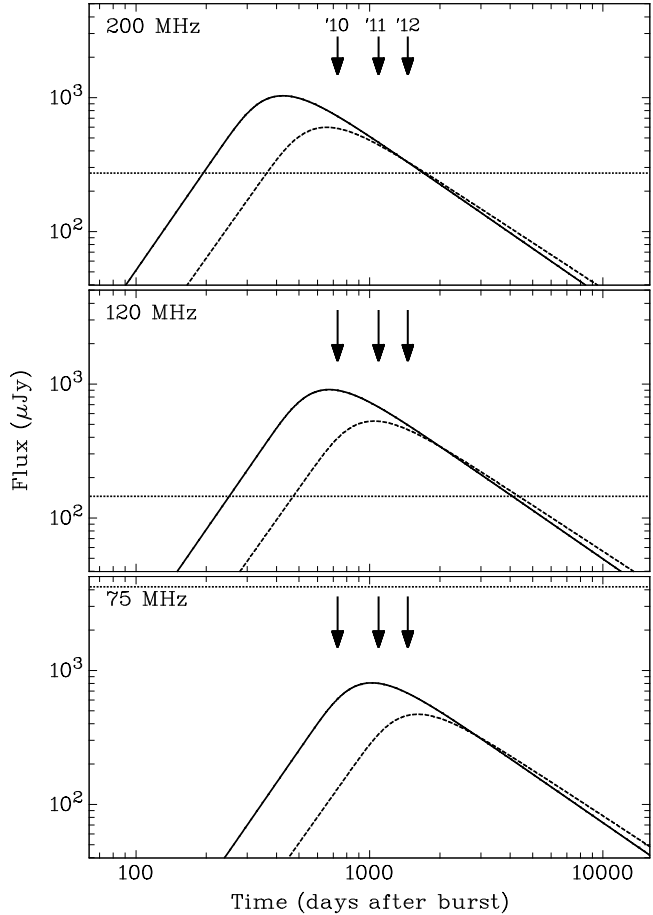}
\caption{Predicted light curves of SN\,2008D (solid lines) and SN\,2007uy (dashed lines) 
at three frequencies in the LOFAR observing range. The dotted lines indicate the LOFAR 
sensitivity, at the 3$\sigma$-level after 4 hours integration. 
The vertical arrows indicate January 1 of the years 2010, 2011 and 2012. } 
\label{fig:SNeLofar}
\end{center}
\end{figure}

Our modeling of the observed broadband radio light curves of SN\,2007uy and SN\,2008D 
enables us to predict the expected light curves in the LOFAR observing range. 
Figure~\ref{fig:SNeLofar} shows the predicted light curves for both sources at three observing frequencies within the LOFAR range, 
two of them in the high band (120-240~MHz) and one in the low band (30-80~MHz). 
The flux of both sources will be above the LOFAR sensitivity limits in the high band, but below the limits in the low band. 
In fact, in Figure~\ref{fig:SNeLofar} one can see that the peak of the spectrum moves through the high band in 2010 
and both SNe will be detectable for the next decade at least. 
We note that future LOFAR observations of these SNe will be affected by refractive ISS, but at the 10$\%$ level at most.

To perform accurate flux measurements it will be sufficient to use all the long baselines in The Netherlands, 
providing a resolution of 2-4~arcseconds, but preferably longer baselines extending into Europe, 
which will provide sub-arcsecond resolution \citep[see e.g.][]{garrett2009arxiv}. 
This kind of resolution will give more accurate flux measurements 
by clearly separating the emission of the SNe from the diffuse host galaxy contribution. 
A follow-up campaign of SNe\,2007uy and 2008D with LOFAR will of course also provide a rich data-set 
on NGC\,2770 at low radio frequencies, a potentially nice extension of the higher frequency radio data obtained with the current radio facilities.

% Conclusions 
\section{Conclusions}\label{sec:conclusions}

We have presented broadband radio observations of the two core-collapse SNe 2008D and 2007uy, and their host galaxy NGC\,2770. 
Detailed modeling of our WSRT and GMRT measurements, spanning $\sim$6 to $\sim$600 days after the SNe explosions 
and observing frequencies ranging from 325~MHz to 8.4~GHz, implies that the radio emission is caused 
by the stellar wind shocked by freely expanding, non-relativistic ejecta as expected in core-collapse SNe. 
We conclude that there is no evidence of a relativistic jet contributing to the observed radio flux for the first year after these two stellar explosions, 
which is further strengthened by our VLBI observations. 
These findings are consistent with the notion that the X-ray outburst from SN\,2008D was not due to a relativistic outflow, 
but was produced by the shock breakout from this SN in a dense stellar wind. 
However, we note that an under-energetic relativistic jet could go undetected and there is still the possibility of the emergence of an off-axis jet. 
Future observations at low radio frequencies, in particular with new-generation radio telescopes like LOFAR, 
will further examine the possible presence of off-axis relativistic jets in these SNe. 

Our CO observations with the ARO 12-m telescope of the host-galaxy show evidence for counter-rotating gas in NGC\,2770, 
only the second galaxy for which this has been found. 
The most plausible explanation for this finding is a collision with the dwarf companion galaxy NGC\,2770B, 
which has led to induced star formation. 
With SN\,2008D and SN\,2007uy occurring within 10 days of each other, 
and the third stripped-envelope core-collapse SN in NGC\,2770 (SN\,1999eh) 9 years earlier, 
we could be witnessing the aftereffects of the collision between these two galaxies.

% Acknowledgements
\acknowledgements{\small We greatly appreciate the support from the WSRT, GMRT, EVN and ARO staff 
in their help with scheduling and obtaining these observations. We thank Bob Campbell and Andreas Brunthaler 
for useful comments regarding astrometric accuracy of the VLBI data. 
The WSRT is operated by ASTRON (Netherlands Institute for Radio Astronomy) 
with support from the Netherlands foundation for Scientific Research. 
The GMRT is operated by the National Center for Radio Astrophysics of the Tata Institute of Fundamental Research. 
The EVN is a joint facility of European, Chinese, South African and other radio astronomy institutes 
funded by their national research councils. 
The National Radio Astronomy Observatory is operated by Associated Universities, Inc., 
under cooperative agreement with the National Science Foundation. 
The ARO 12-m Telescope is operated by the Arizona Radio Observatory, Steward Observatory, University of Arizona. 
AJvdH was supported by an appointment to the NASA Postdoctoral Program 
at the MSFC, administered by Oak Ridge Associated Universities through a contract with NASA. 
APK was supported by NWO-Vici grant C.2320.0017 and also gratefully acknowledges hospitality in January 2009 
provided by the IUCAA where part of this work was carried out. 
ZP acknowledges support from the Hungarian Scientific Research Fund (OTKA, grant K72515). 
JG acknowledges a Royal Society Wolfson Research Merit Award.}

\appendix

\section{Interstellar Scintillation Calculation}\label{sec:appendix}
The type (weak, refractive or diffractive) and strength of the ISS that is affecting the radio fluxes of compact sources 
depends on the observing frequency and on the relative angular sizes of the source and the first Fresnel zone 
\citep{walker1998mnras} of the scattering medium. 
If the source size is smaller than the angular size of the first Fresnel zone, the modulation index, i.e. the fractional flux variation, 
can be rather large, but is quenched significantly when the source size exceeds this characteristic angular scale. 

In Section~\ref{subsec:model} we have modeled the radial expansion rate of the radio ejecta, 
which can be written in terms of the angular source size: $\theta_{\rm{s}}=24.1~t_{10}^{\,0.94}~\mu\rm{as}$ for SN\,2008D 
and $\theta_{\rm{s}}=12.8~t_{10}^{\,0.96}~\mu\rm{as}$ for SN\,2007uy. 
To estimate the angular scales for scintillation, we adopt the \citet{cordes2002} model for the Galactic distribution of free electrons 
and determine the scattering measure to be $SM=2.93\times10^{-4}$~kpc/m$^{(20/3)}$, and the transition frequency $\nu_0$
between weak and strong ISS, $\nu_0=10.42$~GHz. 
The latter indicates that our WSRT and GMRT fluxes are all affected by strong ISS, 
with observing frequencies below $\nu_0$, while the VLA fluxes are affected by weak ISS. 
We note that in the strong ISS regime one has to distinguish between the slow and broadband refractive scintillation, 
and the fast and narrowband diffractive scintillation. 

Based on the scattering measure and transition frequency, we have determined the angular size of the first Fresnel zone 
at the transition frequency: $\theta_{\rm{0}}=0.910~\mu\rm{as}$ \citep{walker1998mnras}. 
In Table~\ref{table:iss} we give the values of the angular scales and modulation indices for the different types of ISS. 
For source sizes smaller than the angular scales give in Table~\ref{table:iss}, 
their respective modulation indices are given by the values in the Table, 
but when the source size becomes larger, the modulation indices are reduced by a factor of 
$(\theta_{\rm{s}}/\theta_{\rm{r}})^{-7/6}$ in the case of weak ISS and $(\theta_{\rm{s}}/\theta_{\rm{d}})^{-1}$ for refractive ISS.

\begin{table}
\begin{center}
\caption{ISS angular scales and modulation indices}
\label{table:iss}
\renewcommand{\arraystretch}{1.25}
\begin{tabular}{|l|l|l|} 
\tableline
ISS type & Angular scale & Modulation index \\
\tableline
Weak         & $\theta_{\rm{w}}=\theta_{\rm{0}}(\nu/\nu_0)^{-1/2}$ & $m_{\rm{w}}=(\nu/\nu_0)^{-17/12}$ \\
                   & \hspace{0.1in}$=2.936~\nu_{\rm{GHz}}^{-1/2}~\mu\rm{as}$ & \hspace{0.1in}$=27.7~\nu_{\rm{GHz}}^{-17/12}$ \\
Refractive & $\theta_{\rm{r}}=\theta_{\rm{0}}(\nu/\nu_0)^{-11/5}$ & $m_{\rm{r}}=(\nu/\nu_0)^{17/30}$ \\
                   & \hspace{0.1in}$=158~\nu_{\rm{GHz}}^{-11/5}~\mu\rm{as}$ & \hspace{0.1in}$=0.265~\nu_{\rm{GHz}}^{17/30}$ \\
Diffractive & $\theta_{\rm{d}}=\theta_{\rm{0}}(\nu/\nu_0)^{6/5}$  & $m_{\rm{d}}=1$ \\
                   & \hspace{0.1in}$=0.0546~\nu_{\rm{GHz}}^{6/5}~\mu\rm{as}$  &  \\
\tableline
\end{tabular}
\end{center}
\end{table}

% Bibliography
\bibliographystyle{aa}
\bibliography{references}

\end{document}